\documentstyle[11pt,moriond,epsfig]{article}

\def\be{\begin{equation}}
\def\ee{\end{equation}}
\def\bea{\begin{eqnarray}}
\def\eea{\end{eqnarray}}

\begin{document}
\vspace*{4cm}
\title{First Results from the PHOBOS Experiment at RHIC}

\author{Christof Roland for the PHOBOS Collaboration \\[0.2cm]
B.B.Back$^1$, M.D.Baker$^2$, 
D.S.Barton$^2$, R.R.Betts$^{6}$, R.Bindel$^7$,  
A.Budzanowski$^3$, W.Busza$^{4}$, \\A.Carroll$^2$,
M.P.Decowski$^4$, 
E.Garcia$^7$, N.George$^1$, K.Gulbrandsen$^4$, 
S.Gushue$^2$, C.Halliwell$^6$, 
G.A.Heintzelman$^2$, C.Henderson$^4$, R.Ho\l y\'{n}ski$^3$, D.J.Hofman$^6$,
B.Holzman$^6$,
E.Johnson$^8$, J.L.Kane$^4$, J.Katzy$^4$, N. Khan$^8$, W.Kucewicz$^{6}$, P.Kulinich$^4$,
W.T.Lin$^5$, S.Manly$^8$,  D.McLeod$^6$, J.Micha\l owski$^3$,
A.C.Mignerey$^7$, J.M\"ulmenst\"adt$^{4}$, R.Nouicer$^6$, 
A.Olszewski$^{2,3}$, R.Pak$^2$, I.C.Park$^8$, 
H.Pernegger$^4$, C.Reed$^4$, L.P.Remsberg$^2$, 
M.Reuter$^6$, C.Roland$^4$, G.Roland$^4$, L.Rosenberg$^4$, 
P.Sarin$^4$, P.Sawicki$^3$, 
W.Skulski$^8$, 
S.G.Steadman$^4$, 
G.S.F.Stephans$^4$, P.Steinberg$^2$, M.Stodulski$^3$, A.Sukhanov$^2$, 
J.-L.Tang$^5$, R.Teng$^8$, A.Trzupek$^3$, 
C.Vale$^4$, G.J.van Nieuwenhuizen$^4$, 
R.Verdier$^4$, B.Wadsworth$^{4}$, F.L.H.Wolfs$^8$, B.Wosiek$^3$, 
K.Wo\'{z}niak$^3$, 
A.H.Wuosmaa$^1$, B.Wys\l ouch$^4$ 
\\[0.2cm]
$^1$ Physics Division, Argonne National Laboratory, Argonne, IL 60439-4843\\
$^2$ Chemistry and C-A Departments, Brookhaven National Laboratory, Upton, NY 11973-5000\\
$^3$ Institute of Nuclear Physics, Krak\'{o}w, Poland\\
$^4$ Laboratory for Nuclear Science, Massachusetts Institute of Technology, Cambridge, MA 02139-4307\\
$^5$ Department of Physics, National Central University, Chung-Li, Taiwan\\
$^6$ Department of Physics, University of Illinois at Chicago, Chicago, IL 60607-7059\\
$^7$ Department of Chemistry, University of Maryland, College Park, MD 20742\\
$^8$ Department of Physics and Astronomy, University of Rochester, Rochester, NY 14627\\[0.4cm]
}

\maketitle\abstracts{
PHOBOS is one of four experiments studying Au-Au collisions at RHIC. During the first
running period RHIC provided Au+Au collisions at $\sqrt{s_{_{NN}}}$ = 56~GeV and 130~GeV. 
The data collected during this period allowed us to study the energy and centrality 
dependence of particle production, the anisotropy of the final state azimuthal distribution 
and particle ratios at mid-rapidity.
}

\section{Introduction}
In June 2000, the Relativistic Heavy Ion Collider (RHIC) at Brookhaven National Laboratory 
delivered the first collisions of gold nuclei at center of mass energies several times 
larger than that previously available at other accelerators. 
The main goal of the RHIC research 
program is to study the behavior of strongly interacting matter under conditions of 
extreme temperatures and energy densities satisfying the prerequisites predicted for 
the creation of the Quark Gluon Plasma (QGP). 

\section{Experimental setup and event selection}
The PHOBOS detector employs multiple arrays of silicon pad detectors to 
perform track reconstruction, vertex detection and multiplicity measurements.
The main components are: a single layer multiplicity detector which covers the 
pseudo-rapidity range $|\eta| < 5.4$ with nearly complete azimuthal coverage, 
a two-layer vertex detector covering $|\eta| < 1.5$ and a magnetic spectrometer with 16 layers
providing momentum measurement and particle identification near mid-rapidity.
Details of the experimental arrangement are described in \cite{phobos1,PAK}. 

The primary event trigger was provided by two sets of 16 scintillator 
paddle counters which detected charged particles in the pseudorapidity range 
$3 < |\eta |< 4.5$. The paddle counters were also used for offline event selection 
and determination of the collision centrality.
Details of this procedure can be found in \cite{judith}.

\begin{figure}[t]
\hspace{1.0cm}
\begin{minipage}{6.5cm}
\epsfig{file=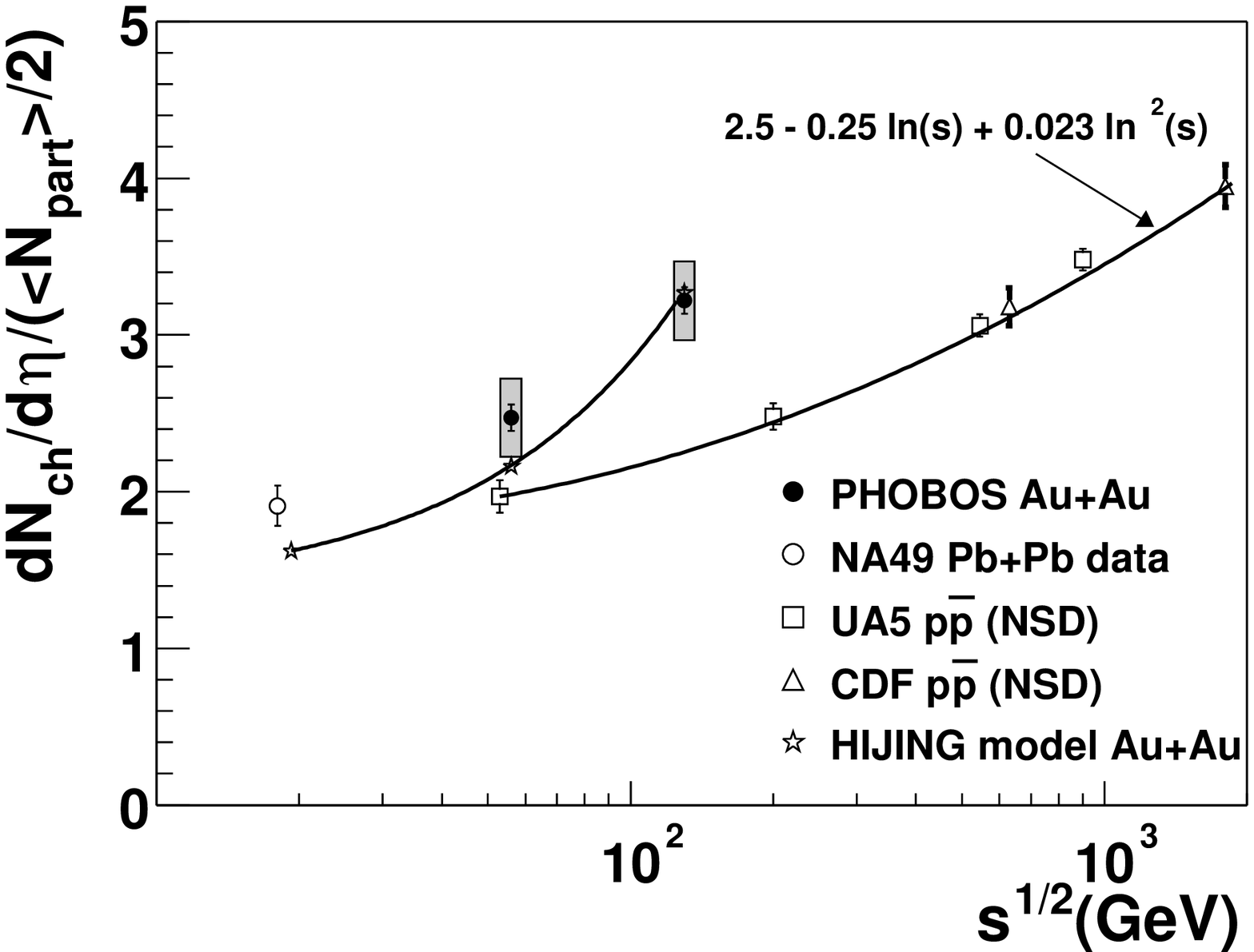,width=6.0cm}
\caption{
Measured pseudorapidity density normalized per
participant pair for central Au+Au collisions. Systematic errors
are shown as shaded areas. 
Also shown are results of Pb+Pb data (CERN SPS), 
HIJING$^{11}$ simulations and a parametrization of $p\overline{p}$ data.
}
\label{dndeta_vs_e}
\end{minipage}
\hspace{1.0cm}
\begin{minipage}{6.5cm}
\epsfig{file=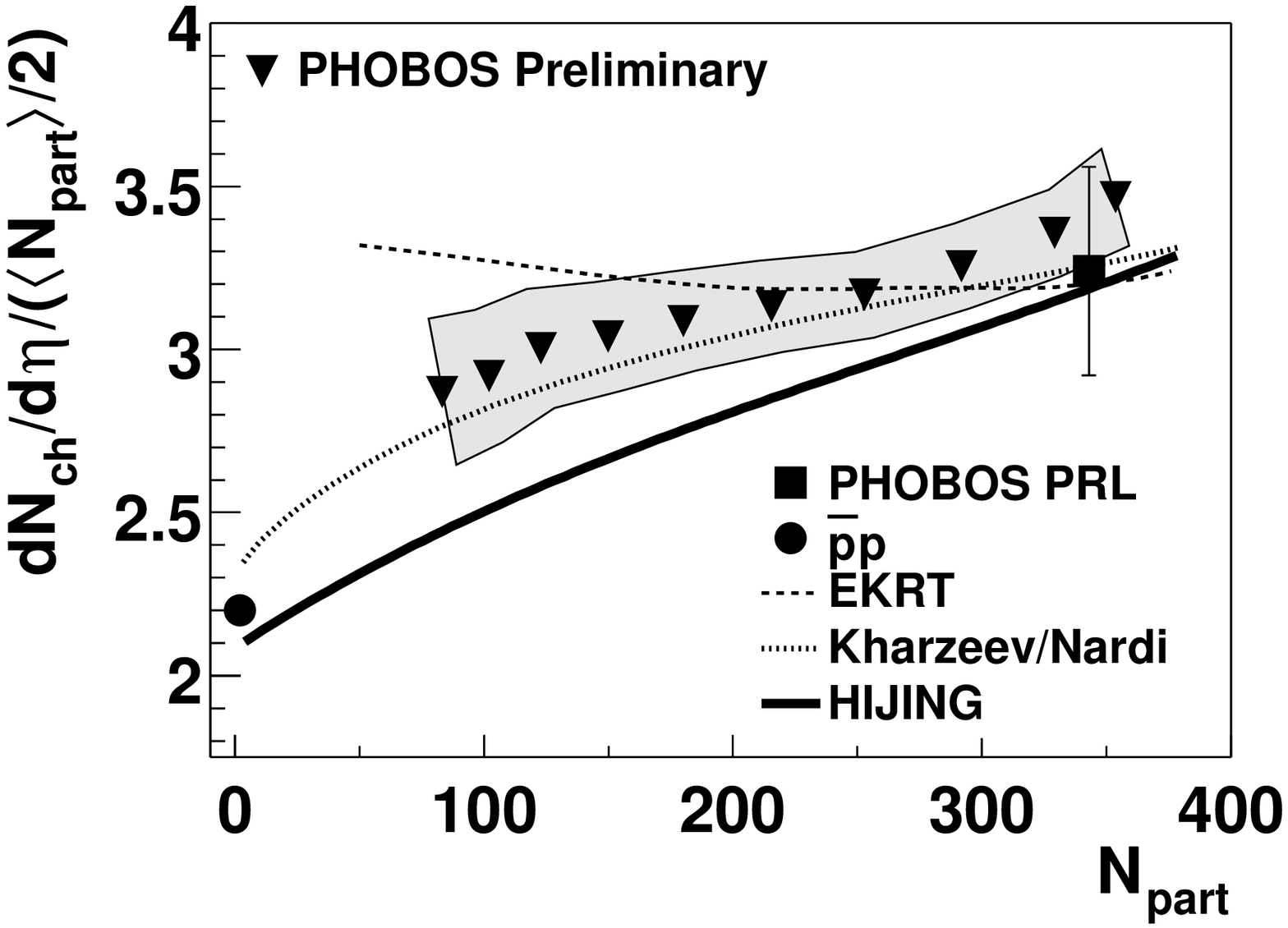,width=6.0cm}
\caption{
Normalized pseudorapidity density 
$dN_{ch}/d\eta|_{|\eta|<1}/(0.5 \times N_{part})$ as a function of the number 
of participants. The solid square is from Ref.$^1$.  Theoretical calculations 
are shown from HIJING$^7$ (solid line), KN$^{8}$ (dotted curve) and EKRT$^{9}$ (dashed curve).
~~~~~~~~~~~~~~~~~~~~~~~}
\label{dndeta_vs_npart}
\end{minipage}
\vspace{-0.5cm}
\end{figure}
\vspace{-0.3cm}
\section{Results}
{\bf Multiplicity and charged particle density.}
The first observable studied was the energy dependence of the 
charged particle pseudo-rapidity density near mid-rapidity for 
central collisions, $dN_{ch}/d\eta |_{|\eta|<1}$.
The measurement is described in detail in \cite{PRL}.
We observed a primary charged particle density of
$dN_{ch}/d\eta |_{|\eta|<1}  = 408 \pm 12 \mbox{(stat)} \pm 30 \mbox{(syst)}$
at $\sqrt{s_{_{NN}}} = 56$~GeV and $555 \pm 12 \mbox{(stat)} \pm 35 \mbox{(syst)}$ 
at $\sqrt{s_{_{NN}}} = 130$~GeV.
Normalizing per participant pair, we deduce
$dN_{ch}/d\eta |_{|\eta|<1} /(0.5 \times N_{part})  = 2.47 \pm 0.1 \mbox{(stat)}
\pm 0.25 \mbox{(syst)}$ and
$3.24 \pm 0.1 \mbox{(stat)} \pm 0.25 \mbox{(syst)}$,
respectively. 
The data were corrected for backgrounds from decay feed-down, secondary 
interactions, stopping particles in the beampipe and detector inefficiencies
using Monte Carlo simulations of our apparatus. 
The normalized yield per participant obtained
for Au+Au collisions, proton-antiproton ($p\overline{p}$) collisions
\cite{pp} and central Pb+Pb collisions \cite{na49} are presented in Fig.~\ref{dndeta_vs_e}. 
Central Au+Au collisions show a significantly larger
charged particle density per participant than non-single 
diffractive (NSD) $p\overline{p}$ collisions at comparable energies.
The observed 31\% increase of density from 56 to 130~GeV in
central Au+Au collisions is significantly larger than the
increase for  $p\overline{p}$ data
in the same energy interval \cite{pp}. 
Comparing our data to those obtained at the
CERN SPS for Pb+Pb collisions at $\sqrt{s_{_{NN}}} = 17.2$~GeV,
we find a 70\% higher charged particle density per participant near $\eta = 0$
at $\sqrt{s_{_{NN}}} = 130$~GeV.
The dependence of the charged particle density $dN_{ch}/d\eta |_{|\eta|<1}$ on
event centrality together with theoretical predictions \cite{hijing-cent,kn,ekrt}
are shown in Fig.~\ref{dndeta_vs_npart}.

Further information on
the underlying physics can be obtained by studying the change 
in particle production over a wider range in $\eta$.
We performed this measurement using the PHOBOS multiplicity detector, 
consisting of the Octagon barrel covering $|\eta |\leq$3.2 and the 
Ring counters detecting particles with $3 \leq |\eta| \leq 5.4$. 
Details of the hit-counting procedure and the background correction
can be found in \cite{alan_qm2001}.
For the most central 3\% of the collisions we find an average number of charged
particles of $4100 \pm 100 \mbox{(stat)} \pm 400 \mbox{(syst)}$ within $|\eta| < 5.4$.

\begin{figure}
\begin{center}
\epsfig{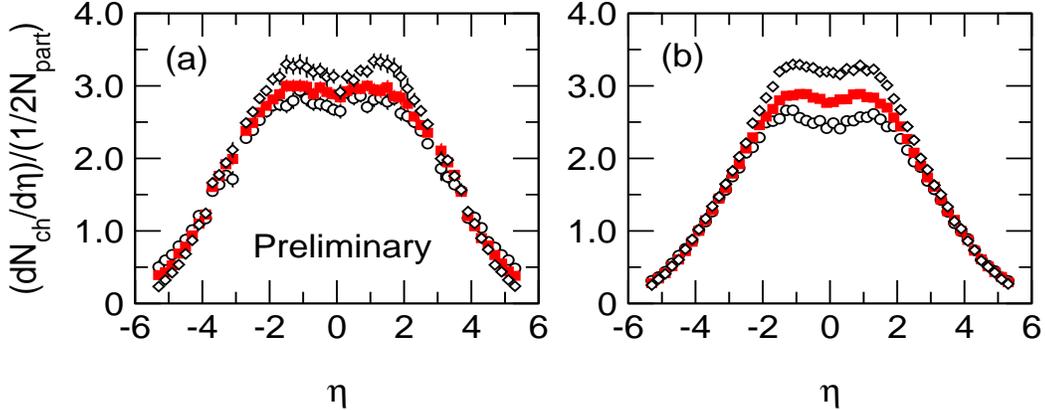}
\end{center}
\vspace{-0.5cm}
\caption{(a). Measured $dN_{ch}/d\eta/(\langle N_{part}\rangle/2)$ for $\langle N_{part} \rangle$=102 (circles), 
216 (squares) and 354 (diamonds). We estimate a systematic uncertainty of 10\% 
independent of $\eta$.
(b). Same as (a) from HIJING.~~~~~~~~~~~~~~~~~~~~~~~~~~~~~~
}
\label{fig2}
\vspace{-0.4cm}
\end{figure}

The measured $dN_{ch}/d\eta$ distributions scaled by $0.5\times N_{part}$ for
three selected centrality bins are shown in Fig.~\ref{fig2}a.
For comparison, the predictions of HIJING \cite{hijing}
appear in Fig.~\ref{fig2}b.  At mid\-rapidity, the non-central data are
higher than the predictions. Both data and HIJING show that the bulk of 
additional particle production in
central collisions, as compared to peripheral ones, occurs near mid-rapidity. 
The central plateau observed in the data is significantly wider than the HIJING prediction.\\

\noindent {\bf Event Anisotropy.}
The momentum space anisotropy, which is a reflection of the initial-state 
space anisotropy for non-central collisions, is commonly quantified by the first 
and second Fourier coefficients of the azimuthal particle distribution
relative to the event-by-event reaction plane,
 ${dN / d(\phi-\Psi_R)}= N_0(1+\sum_i 2v_i \cos[i(\phi-\Psi_R)])$, 
where $\Psi_R$ is the reaction plane angle. Usually, the 
first Fourier component ($v_1$) is referred to as directed flow and 
the second one ($v_2$) as elliptic flow. 

In PHOBOS, the elliptic flow was calculated from the azimuthal distribution
of hits in the Octagon detector. For every event the event plane was 
determined by a fit to the azimuthal hit distribution. 
The observed raw value of $v_2$ was corrected for the effects of detector 
granularity, the resolution of the event-by-event determination of the 
event plane relative to the true reaction plane and the contribution of background hits.
Details of the analysis can be found in \cite{inkyu_qm2001}.

\begin{figure}[t]
\begin{minipage}{7.5cm}
\mbox{\epsfig{file=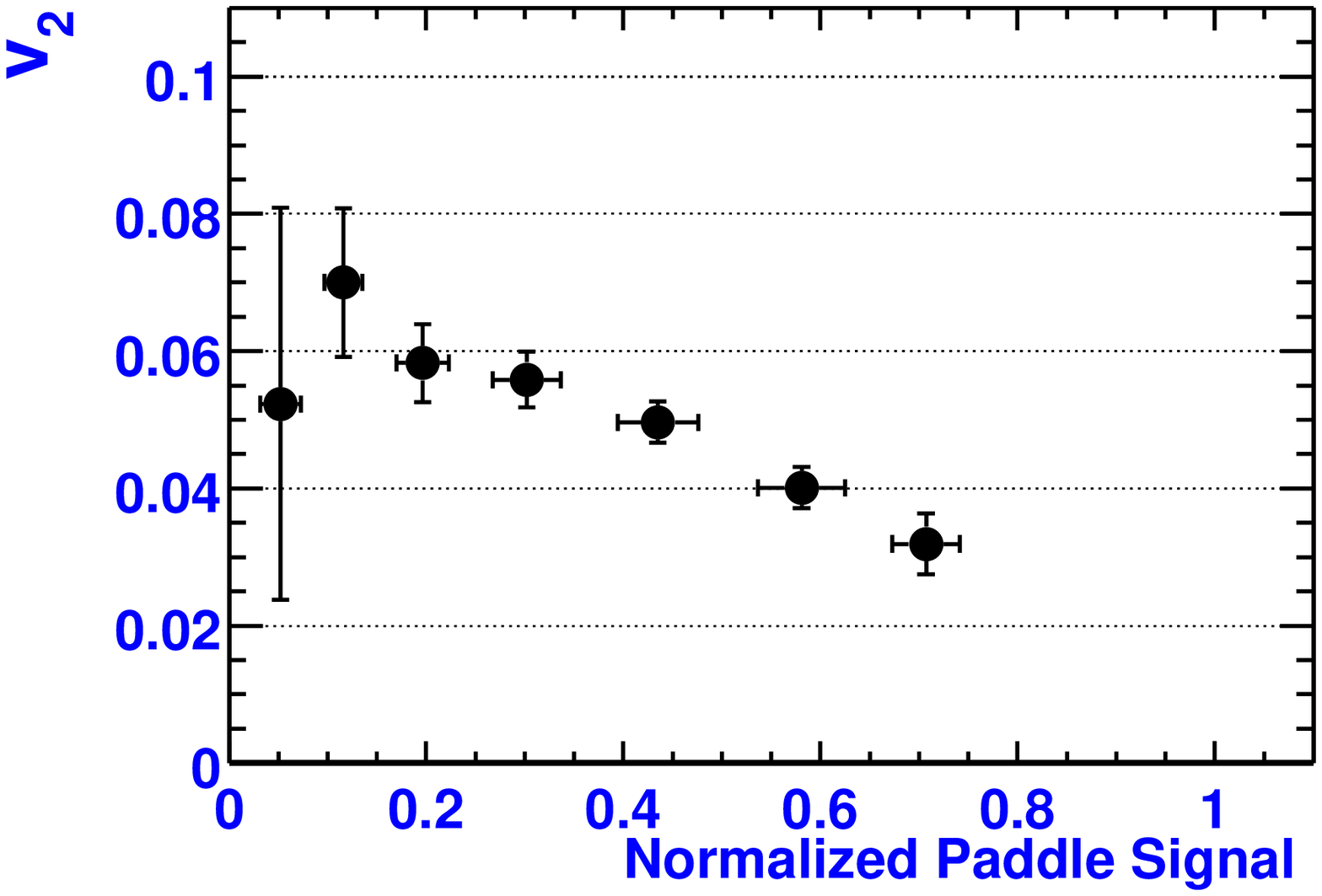,width=7cm}}
\put(-160,110){\mbox{\tiny \bf PHOBOS Preliminary}}
\caption{Elliptic Flow $v_2$, in the region -1.0$<\eta<$1.0 as a function of normalized 
paddle signal. The error bars are statistical only. The estimated systematic error 
is $\Delta v_2 = 0.007$.~~~~~~~~~~~~~~~~~~~~~~~~~~~~~~~~~~~~~~~~~~ }
\label{v2_vs_npart}
\end{minipage}
\hspace{0.5cm}
\begin{minipage}{7.5cm}
\mbox{\epsfig{file=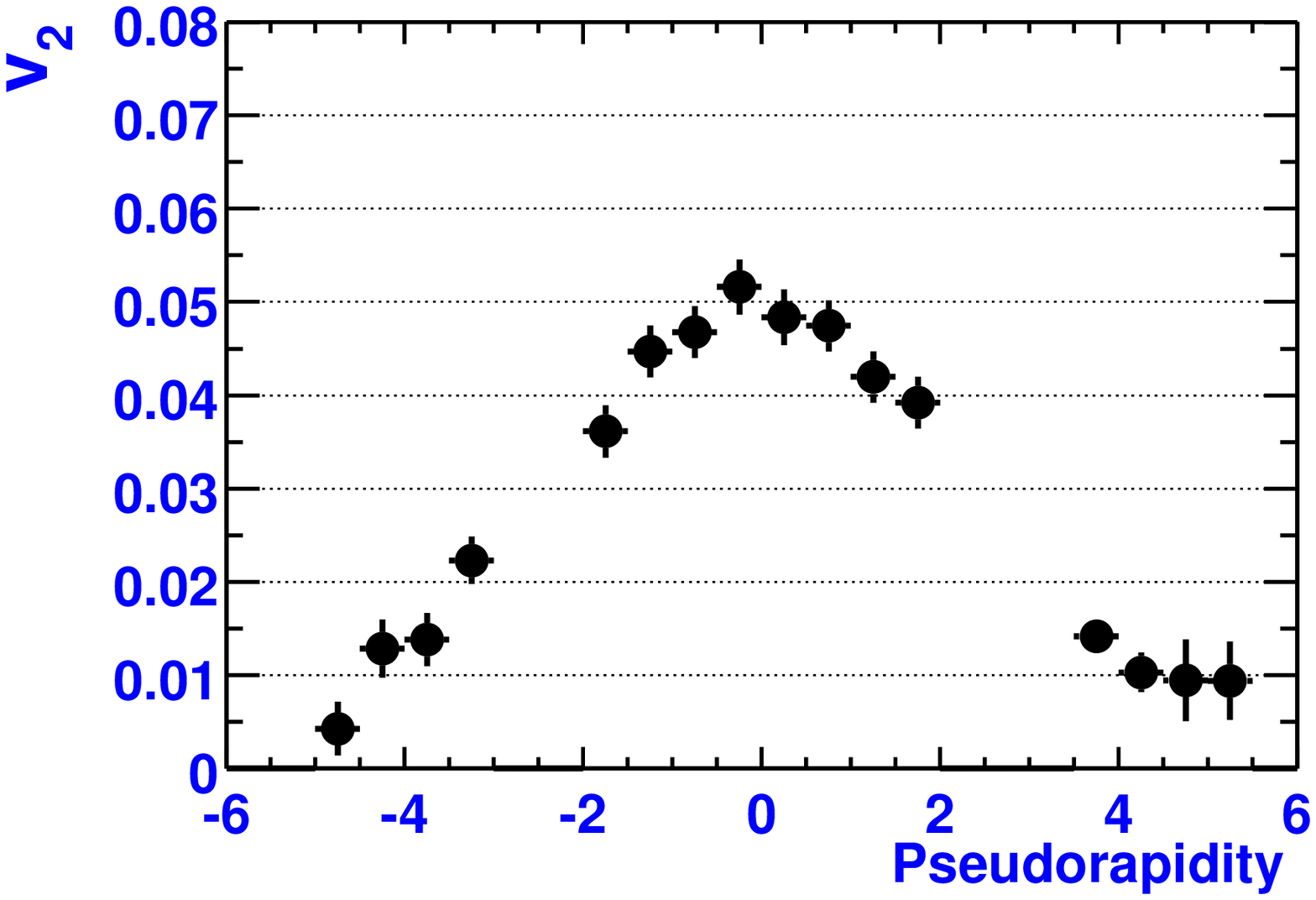,width=7cm}} 
\put(-160,110){\mbox{\tiny \bf PHOBOS Preliminary}}
\caption{Elliptic Flow $v_2$, averaged over centrality, as a function of
pseudorapidity $\eta$. The error bars are statistical only. The estimated systematic error is $\Delta v_2 = 0.007$.~~~~~~~~~~~~~~~~~~~~~~~~~~~~~~~~~~~~~~~~~~~~~~~~~~~~~ } 
\label{v2_vs_eta}
\end{minipage}
\vspace{-0.8cm}
\end{figure}

We determined $v_2$ for events in seven classes of centrality based on the 
normalized paddle signal. 
We observed a strong centrality dependence of $v_2$ ranging from 0.03 for the most 
central events up to 0.07 for peripheral events (Fig.~\ref{v2_vs_npart}). 
We have also studied the pseudorapidity dependence of $v_2$ over a large range 
in centrality (Fig.~\ref{v2_vs_eta}). Outside of the central pseudorapidity 
region, i.e. $|\eta| > 1$, $v_2$ drops very quickly, approximately following the 
shape of the charged particle distribution (see Fig.~\ref{fig2}).\\

\noindent {\bf Particle ratios.}
The ratio of antibaryons to baryons plays a key role in understanding
the properties of the hot and dense system formed in heavy-ion 
collisions at high energies. 
Using the mid-rapidity spectrometer, we have studied the ratios of 
multiplicities of particles and antiparticles for primary charged pions, kaons 
and protons at $\sqrt{s_{_{NN}}} = 130$~GeV. 

The particle ratios were measured for the 12\% most central 
events, for which  we estimate $\langle N_{part} \rangle = 312 \pm 10 \mbox{(syst)}$.
As the geometrical layout of the PHOBOS detector leads to an asymmetry 
in the acceptance and detection efficiency for positively and 
negatively charged particles in the same events, we used data taken with both 
polarities of the PHOBOS magnet.
The most important systematic effects such as geometrical acceptance and tracking 
efficiency cancel in the ratio. 
The raw particle ratios are corrected for particle decays, secondary 
interactions and the contribution from
feed-down of strange hadrons. Further details can be found in \cite{nigel_qm2001}.
Within our acceptance we find the following ratios:

\hspace{2.7cm}
\begin{minipage}{1.5cm}
\begin{eqnarray*}
&\langle \pi^- \rangle / \langle \pi^+ \rangle & =  1.00 \pm 0.01\mbox{(stat)}\pm 0.02\mbox{(syst)}\\
&\langle K^- \rangle/ \langle K^+ \rangle & =  0.91\pm 0.07\mbox{(stat)}\pm 0.06\mbox{(syst)}\\
&\langle \overline{p} \rangle / \langle p \rangle & =  0.60\pm 0.04\mbox{(stat)}\pm 0.06\mbox{(syst)}\\
\end{eqnarray*}
\end{minipage}

We estimated a baryo-chemical potential $\mu_B$ from the
$\langle K^- \rangle/ \langle K^+ \rangle$ and
$\langle \overline{p} \rangle / \langle p \rangle$ ratios, using a
statistical model calculation \cite{redlich_qm2001}. 
For a realistic range of freeze-out temperatures of 160 to 170 MeV, both ratios
are consistent with $\mu_B  = 45 \pm 5$~MeV.
This value is much lower than $\mu_B = 240-270$~MeV obtained in thermal
fits to Pb+Pb data from the CERN SPS \cite{becattini,pbm}, showing a closer but
not yet complete approach to a baryon-free regime at RHIC.

\vspace{-0.3cm}

\section*{Acknowledgments}
\vspace{-0.3cm}
This work was partially supported by: (US) DoE grants DE-AC02-98CH10886,
DE-FG02-93ER40802, DE-FC02-94ER40818, DE-FG02-94ER40865,
DE-FG02-99ER41099 and W-31-109-ENG-38,
NSF grants 9603486, 9722606 and 0072204,
(Poland) KBN grant 2 P03B 04916,
(Taiwan) NSC contract NSC 89-2112-M-008-024.

\end{document}